\documentclass[aip,apl,twocolumn,reprint]{revtex4-1}
\usepackage{graphicx}
\usepackage{dcolumn}
\usepackage{bm}
\usepackage{color}
\usepackage{amsmath}
\usepackage{accents}

\begin{document}


\title{Large spin pumping effect in antisymmetric precession of Ni$_{79}$Fe$_{21}$/Ru/Ni$_{79}$Fe$_{21}$}

\author{ H. Yang}
\author{Y. Li}
\author{W.E. Bailey}\email{Contact author.  web54@columbia.edu}

\affiliation{Materials Science and Engineering, Dept. of Applied Physics and Applied Mathematics, Columbia University, New York NY 10027}
\date{\today}

\begin{abstract}

In magnetic trilayer structures, a contribution to the Gilbert damping of ferromagnetic resonance arises from spin currents pumped from one layer to another.  This contribution has been demonstrated for layers with weakly coupled, separated resonances, where magnetization dynamics are excited predominantly in one layer and the other layer acts as a spin sink.  Here we show that trilayer structures in which magnetizations are excited simultaneously, antisymmetrically, show a spin-pumping effect roughly twice as large.  The antisymmetric (optical) mode of antiferromagnetically coupled Ni$_{79}$Fe$_{21}$(8nm)/Ru/Ni$_{79}$Fe$_{21}$(8nm) trilayers shows a Gilbert damping constant greater than that of the symmetric (acoustic) mode by an amount as large as the intrinsic damping of Py ($\Delta \alpha\simeq\textrm{0.006}$).  The effect is shown equally in field-normal and field-parallel to film plane geometries over 3-25 GHz.  The results confirm a prediction of the spin pumping model and have implications for the use of synthetic antiferromagnets (SAF)-structures in GHz devices.

\end{abstract}

\maketitle

Pumped spin currents\cite{tserkovnyakPRL02,tserkovnyakRMP05} are widely understood to influence the magnetization dynamics of ultrathin films and heterostructures.  These spin currents increase the Gilbert damping or decrease the relaxation time for thin ferromagnets at GHz frequencies.  The size of the effect has been parametrized through the effective spin mixing conductance $g_r^{\uparrow\downarrow}$, which relates the spin current pumped out of the ferromagnet, transverse to its static (time-averaged) magnetization, to its precessional amplitude and frequency.  The spin mixing conductance is interesting also because it determines the transport of pure spin current across interfaces in quasistatic spin transport, manifested in e.g. the spin Hall effect.

In the spin pumping effect, spin current is pumped away from a ferromagnet / normal metal (F$_{1}$/N) interface, through precession of $F_1$, and is absorbed elsewhere in the structure, causing angular momentum loss and damping of $F_1$.  The spin current can be absorbed through different processes in different materials.  When injected into paramagnetic metals (Pt, Pd, Ru, and others), the spin current relaxes exponentially with paramagnetic layer thickness\cite{mizukamiPRB02, ghoshAPL11,ghoshPRL12}.  The relaxation process has been likened to spin-flip scattering as measured in CPP-GMR, where spin-flip events are localized to heavy-metal impurities\cite{bassJPCM07} and the measurement reveals the spin diffusion length $\lambda_{SD}$.  When injected into other ferromagnets ($F_2$ in F$_{1}$/N/F$_{2}$), the spin current is absorbed through its torque on magnetization\cite{woltersdorfPRL07,ghoshPRL12}.  A similar process appears to be relevant for antiferromagnets as well\cite{merodioAPL14}.

In F$_{1}$/N/F$_{2}$ structures, only half of the total possible spin pumping effect has been detected up until now.  For well-separated resonances of $F_1$ and $F_2$, only one layer will precess with large amplitude at a given frequency $\omega$, and spin current is pumped from a precessing $F_1$ into a static $F_2$.  If both layers precess symmetrically, with the same amplitude and phase, equal and opposite spin currents are pumped into and out of each layer, causing no net effect on damping.  The difference between the symmetric mode and the uncoupled mode, increased by a spin pumping damping $\alpha_{sp}$ was detected first in epitaxial Fe/Au/Fe structures\cite{heinrichPRL03}.  However, if the magnetizations can be excited with {\it antisymmetric} precession, the coupled mode should be damped by twice that amount, $2\alpha_{sp}$.  Takahashi\cite{takahashiAPL15} has published an explicit prediction of this "giant spin pumping effect" very recently, including an estimate of a fourfold enhanced spin accumulation in the central layer.

In this paper, we show that a very large spin pumping effect can be realized in antisymmetric precession of Py(8 nm)/Ru(0.70-1.25 nm)/Py(8 nm) synthetic antiferromagnets (SAF, Py=Ni$_{79}$Fe$_{21}$).  The effect is roughly twice that measured in Py trilayers with uncoupled precession.  Variable-frequency ferromagnetic resonance (FMR) measurements show, for structures with magnetization saturated in the film plane or normal to the film plane, that symmetric (acoustic mode) precession of the trilayer has almost no additional damping, but the antisymmetric (optical mode) precession has an additional Gilbert damping of $\sim\textrm{0.006}$, compared with an uncoupled Py(8nm) layer in a F$_{1}$/N/F$_{2}$ structure of $\sim\textrm{0.003}$.  The interaction stabilizes the antiparallel magnetization state of SAF structures, used widely in different elements of high-speed magnetic information storage, at GHz frequencies.

{\it Method:} Ta(5 nm)/Cu(3 nm)/Ni$_{79}$Fe$_{21}$(8 nm)/Ru($t_{Ru}$)/Ni$_{79}$Fe$_{21}$(8 nm)/Cu(3 nm)/SiO$_{2}$(5 nm), $t_{Ru}=\textrm{0.7 - 1.2 nm}$  heterostructures were deposited by ultrahigh vacuum (UHV) sputtering at a base pressure of 5 $\times$ 10$^{-9}$ Torr on thermally oxidized Si substrates.  The Ru thckness range was centered about the second antiferromagnetic maximum of interlayer exchange coupling (IEC) for Py/Ru/Py superlattices, 8-12 \AA	, established first by Brillouin light scattering (BLS) measurement\cite{fassbenderJMMM93}.  Oscillatory IEC in this system, as a function of $t_{Ru}$, is identical to that in the more widely studied Co/Ru($t_{Ru}$)/Co superlattices\cite{wigenPRB94}, 11.5 \AA, but is roughly antiphase to it.  An in-plane magnetic field bias of 200 G, rotating in phase with the sample, was applied during deposition as described in \cite{chengRSI12}.  

The films were characterized using variable frequency, swept-field, magnetic-field modulated ferromagnetic resonance (FMR).  Transmission measurements were recorded through a coplanar waveguide (CPW) with center conductor width of 300 $\mu$m, with the films placed directly over the center conductor, using a microwave diode signal locked in to magnetic field bias modulation.  FMR measurements were recorded for magnetic field bias $H_B$ applied both in the film plane (parallel condition, $pc$) and perpendicular to the plane (normal condition, $nc$.)  An azimuthal alignment step was important for the $nc$ measurements.  For these, the sample was rotated on two axes to maximize the field for resonance at 3 GHz.  

For all FMR measurements, the sample magnetization was saturated along the applied field direction, simplifying extraction of Gilbert damping $\alpha$.  The measurements differ in this sense from low-frequency measurements of similar Py/Ru/Py trilayer structures by Belmenguenai et al\cite{belmenguenaiPRB07}, or broadband measurements of (stiffer) [Co/Cu]$_{\times\textrm{10}}$  multilayers by Tanaka et al\cite{tanakaAPEX14}.  In these studies, effects on $\alpha$ could not be distinguished from those on inhomogeneous broadening.  

{\it Model:} In the measurements, we compare the magnitude of the damping, estimated by variable-frequency FMR linewidth through $\Delta H_{1/2} = \Delta H_0 + 2\alpha\omega/|\gamma|$, and the interlayer exchange coupling (IEC) measured through the splitting of the resonances.  Coupling terms between layers $i$ and $j$ are introduced into the Landau-Lifshitz-Gilbert equations for magnetization dynamics through

\begin{widetext}
\begin{equation}
\dot{\mathbf{m_i}} = -\mathbf{m_i}\times\left(\gamma_i\mathbf{H_{eff}}+\omega_{ex,i}\mathbf{m_j}\right)+\alpha_0\mathbf{m_i}\times\dot{\mathbf{m_i}}+\alpha_{sp,i}\left(\mathbf{m_i}\times\mathbf{\dot{m_i}}-\mathbf{m_j}\times\mathbf{\dot{m_j}}\right) 
\label{eq_llg}
\end{equation}
\end{widetext}

in $cgs$ units, where we defined magnetization unit vectors as $\mathbf{m_1}=\mathbf{M_1}/M_{s,1}$, $\mathbf{m_2}=\mathbf{M_2}/M_{s,2}$ with $M_{s,i}$ the saturation moments of layer $i$.  The coupling constants are, for the IEC term, $\omega_{ex,i}\equiv \gamma_i A_{ex}/(M_{s,i}\: t_{i})$, where the energy per unit area of the system can be written $u_{A}=-A_{ex}\mathbf{m_i}\cdot\mathbf{m_j}$, and $t_i$ is the thickness of layer $i$.  Positive values of $A_{ex}$ correspond to ferromagnetic coupling, negative values to antiferromagnetic coupling.  The spin pumping damping term is $\alpha_{sp,i}\equiv \gamma\hbar \tilde{g}_{\uparrow\downarrow}^{FNF}/( 4\pi M_{s,i} t_{i})$, where $\tilde{g}_{\uparrow\downarrow}^{FNF}$ is the spin mixing conductance of the trilayer in units of nm$^{-2}$.  $\alpha_0$ is the bulk damping for the layer.

\begin{figure}
  \includegraphics[width=\columnwidth]{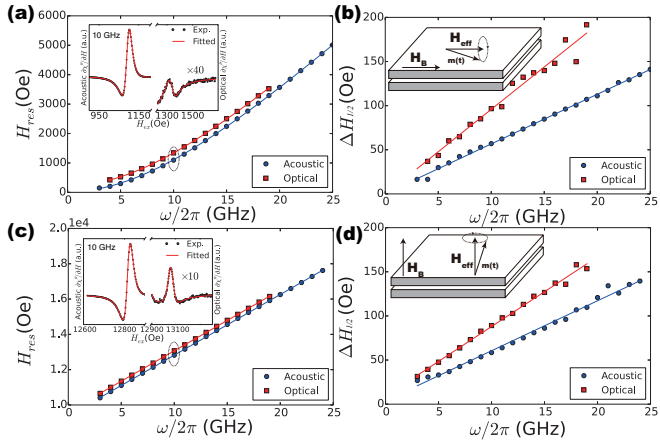}
  \caption{FMR measurement of Ni$_{79}$Fe$_{21}$(8 nm)/Ru($t_{Ru}$)/Ni$_{79}$Fe$_{21}$(8 nm) trilayers; example shown for $t_{Ru}=\textrm{1.2 nm}$.  {\it Inset}: lock-in signal, transmitted power at 10 GHz, as a function of bias field $H_B$, for a) {\it pc-}FMR and c) {\it nc-}FMR.  A strong resonance is observed at lower $H_B$ and a weaker one at higher $H_B$, attributed to the symmetric (S) and antisymmetric (A) modes, respectively.  a), c): Field for resonance $\omega(H_B)$ for the two resonances.  Lines are fits to the Kittel resonance expression, assuming an additional, constant, positive field shift for $\omega_A$, $H_{ex}=-2 A_{ex}/(M_s t_F)$ due to antiferromagnetic interlayer coupling $A_{ex}<0$.  b) {\it pc-FMR} and d) {\it nc}-FMR linewidth as a function of frequency $\Delta H_{pp}(\omega)$ for fits to Gilbert damping $\alpha$.}
  \label{fig0}
\end{figure}

The collective modes of $1,2$ are found from small-amplitude solutions of Equations \ref{eq_llg} for $i=1,2$.  General solutions for resonance frequencies with arbitrary magnetization alignment, not cognizant of any spin pumping damping or dynamic coupling, were developed by Zhang et al\cite{wigenPRB94}.  In our experiment, to the extent possible, layers $1,2$ are identical in deposited thickness, magnetization, and interface anisotropy (each with Cu the opposite side from Ru).  Therefore if $\omega_i$ represents the FMR frequency (dependent on bias field $H_B$) of each layer $i$, the two layers have $\omega_1=\omega_2=\omega_0$.  In this limit, there are two collective modes: a perfectly symmetric mode $S$ and a perfectly antisymmetric mode $A$ with complex frequencies $\widetilde{\omega_S}=(1-i\alpha_0)\omega_0$ and $\widetilde{\omega_A}=(1-i\alpha_0-2i\alpha_{sp})\left(\omega_0+2\omega_{ex}\right)$.  The Gilbert damping for the modes, $\alpha_k=-\textrm{Im}(\widetilde{\omega_{k}})/\textrm{Re}(\undertilde{\omega_{k}})$, where $k=(S,A)$, and the resonance fields $H_B^k$ satisfy

\begin{eqnarray}
 H_B^A &=& H_B^S +2 H_{ex}\qquad H_{ex}= -A_{ex}/(M_s t_F) \label{delta_res} \\
 \alpha_A &=& \alpha_S + 2\alpha_{sp} \qquad \alpha_{sp}= \gamma\hbar \tilde{g}_{\uparrow\downarrow}^{FNF}/( 4\pi M_{s,i} t_{i})\label{delta_alpha}
\end{eqnarray}

and $\omega_{ex}=\gamma H_{ex}$.  Note that there is no relationship in this limit between the strength of the exchange coupling $A_{ex}$ and the spin-pumping damping $2\alpha_{sp}$ expressed in the antisymmetric mode.  The spin pumping damping and the interlayer exchange coupling can be read simply from the differences in the the Gilbert damping $\alpha$ and resonance fields between the antisymmetric and symmetric modes.  The asymmetric mode will have a higher damping by $2\alpha_{sp}$ for any $A_{ex}$ and a higher resonance field for $A_{ex}<0$, i.e. for antiferromagnetic IEC: because the ground state of the magnetization is antiparallel at zero applied field, antisymmetric excitation rotates magnetizations towards the ground state and is lower in frequency than symmetric excitation.  

{\it Results:}  Sample $pc-$ and $nc-$FMR data are shown in Figure \ref{fig0}.  Raw data traces (lock-in voltage) as a function of applied bias field $H_B$ at 10 GHz are shown in the inset.  We observe an intense resonance at low field and resonance weaker by a factor of 20-100 at higher field.  On the basis of the intensities, as well as supporting MOKE measurements, we assign the lower-field resonance to the symmetric, or "acoustic" mode and the higher-field resonance to the antisymmetric, or "optical" mode.  Similar behavior is seen in the $nc$- and $pc-$FMR measurements. 

In Figure \ref{fig0}a) and c), which summarizes the fields-for-resonance $\omega(H_B)$, there is a rigid shift of the antisymmetric-mode resonances to higher bias fields $H_B$, as predicted by the theory.  The lines show fits to the Kittel resonance, $\omega_{pc}=\gamma\sqrt{H_{eff}\left(H_{eff}+4\pi M_s^{eff}\right)}$, $\omega_{nc}=\gamma\left(H_{eff}-4\pi M_s^{eff}\right)$ with an additional effective field along the magnetization direction for the antisymmetric mode: $H_{eff,S}=H_B$, and  $H_{eff,A} = H_B-8\pi A_{ex}/(4\pi M_s t_F)$.

\begin{figure}
  \includegraphics[width=\columnwidth]{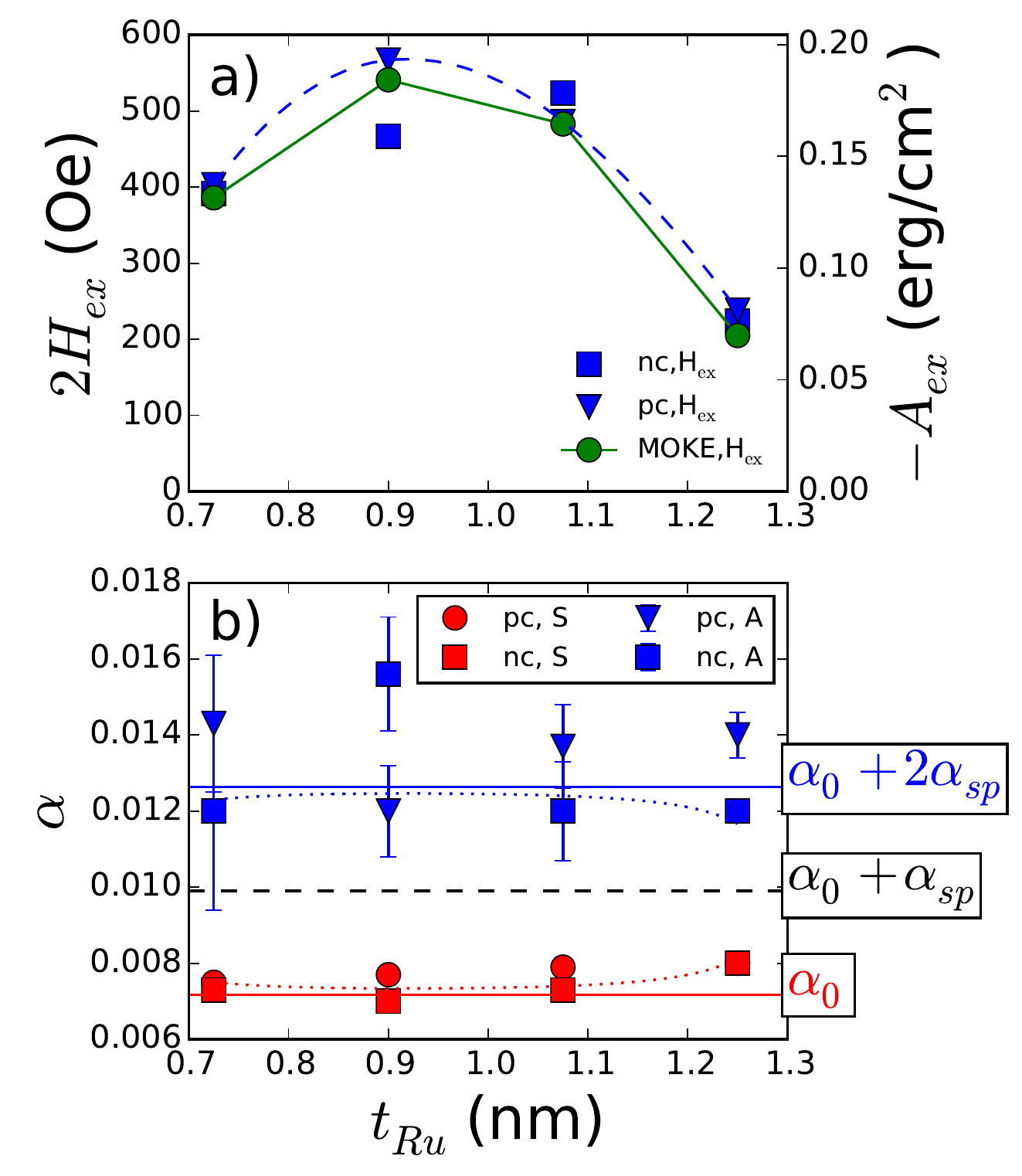}
  \caption{Coupling parameters for Py/Ru/Py trilayers.  a):   {\it Interlayer (static) coupling} from resonance field shift of antisymmetric mode; see Fig \ref{fig0} a),c).  The antiferromagnetic exchange parameter $A_{ex}$ is extracted through Eq \ref{delta_res}, in agreement with values found in Ref \cite{fassbenderJMMM93}.  The line is a guide to the eye.  b) {\it Spin pumping (dynamic) coupling} from damping of the symmetric (S) and antisymmetric (A) modes; see Fig \ref{fig0} b), d).  The spin pumping damping for uncoupled layer precession in Py/Ru/Py, $\alpha_{sp}$ is shown for comparsion.  Dotted lines show the possible effect of $\sim$100 Oe detuning for the two Py layers.  See text for details.}\label{fig1}
\end{figure}

In Figure \ref{fig1}, we show coupling parameters, as a function of Ru thickness, extracted from the FMR measurements illustrated in Figure \ref{fig0}.  Coupling fields are measured directly from the difference between the symmetric and antisymmetric mode positions and plotted in Figure \ref{fig1}a.  We convert the field shift to antiferromagnetic IEC constant $A_{ex}<0$ through Equation \ref{delta_res}, using the thickness $t_{F}=\textrm{8 nm}$ and bulk magnetization $4\pi M_s=\textrm{10.7 kG}$\cite{ghoshAPL11}.  The extracted exchange coupling strength in {\it pc}-FMR has a maximum antiferromagnetic value of $A_{ex}=-\textrm{0.2 erg/cm}^2$, which agrees to 5\% with that measured by Fassbender et al\cite{fassbenderJMMM93} for [Py/Ru]$_N$ superlattices.  

The central result of the paper is shown in Figure \ref{fig1} b).  We compare the damping $\alpha$ of the symmetric ($S$) and antisymmetric ($A$) modes, measured both through {\it pc-}FMR and {\it nc}-FMR.  The values measured in the two FMR geometries agree closely for the symmetric modes, for which signals are larger and resolution is higher.  They agree roughly within experimental error for the antisymmetric modes, with no systematic difference.  The antisymmetric modes clearly have a higher damping than the symmetric modes.  Averaged over all thickness points, the enhanced damping is roughly $\alpha_{A}-\alpha_S=\textrm{0.006}$.  

{\it Discussion:}  The damping enhancement of the antisymmetric ($A-$) mode over the symmetric ($S-$) mode, shown in Figure \ref{fig1}b), is a large effect.  The value is close to the intrinsic bulk damping $\alpha_0\sim\textrm{0.007}$ for Ni$_{79}$Fe$_{21}$.  We compare the value with the value $2\alpha_{sp}$ expected from theory for the antisymmetric mode and written in Eq \ref{delta_alpha}.  The interfacial spin mixing conductance for Ni$_{79}$Fe$_{21}$/Ru, was found in Ref. \cite{beheraJAP15} to be $\tilde{g}_{\uparrow\downarrow}^{FN}=\textrm{24 nm}^{-2}$.  For a F/N/F structure, in the limit of ballistic transport with no spin relaxation through $N$, the effective spin mixing conductance is $\tilde{g}_{\uparrow\downarrow}^{FN}/2$: spin current must cross two interfaces to relax in the opposite $F$ layer, and the conductance reflects two series resistances\footnote{see eqs. 31, 74, 81 in Ref \cite{tserkovnyakRMP05}}.  This yields $\alpha_{sp}=0.0027$.  The observed enhancement matches well with, and perhaps exceeds slightly, the "giant" spin pumping effect of $2\alpha_{sp}$, as shown.	

Little dependence of the Gilbert damping enhancement $\alpha_A-\alpha_S$ on the resonance field shift $H_A-H_S$ can be observed in Figure \ref{fig1} a,b.  We believe that this independence reflects close tuning of the resonance frequencies for Py layers 1 and 2, as designed in the depositions.  For finite detuning $\Delta\omega$ defined through $\omega_2 = \omega_0+\Delta\omega$ and $\omega_1=\omega_0-\Delta\omega$, the modes change.  Symmetric and antisymmetric modes become hybridized as $S^{'}$ and $A^{'}$, and the difference in damping is reduced.  Defining $\widetilde{\Delta\omega}^2=\left(1-i\alpha_0\Omega\right)\left(1-i\alpha_0\Omega-2i\alpha_{sp}\Omega\right)\Delta\omega^2$, it is straightforward to show that for the {\it nc-}case, the mode frequencies are $\omega_{S',A'} = \left(\widetilde{\omega_S}+\widetilde{\omega_A}\right)/2 \pm \sqrt{\left(\widetilde{\omega_S}-\widetilde{\omega_A}\right)^2/4+\widetilde{\Delta\omega\:^2}}$.  The relevant parameter is the frequency detuning normalized to the exchange (coupling) frequency, $z\equiv\Delta\omega/(2\omega_{ex})$; if $z\gg 1$, the layers have well-separated modes, and each recovers the {\it uncoupled} damping enhancement of $\alpha_{sp}$, $\alpha_{S',A'}=\alpha_0 + \alpha_{sp}$ identified in Refs \cite{heinrichPRL03,ghoshPRL12}.

The possibility of finite detuning, assuming $(\omega_2-\omega_1)/\gamma = \textrm{100 Oe}$, is shown in Figure \ref{fig1}b), with the dotted lines.  The small$-z$ limit for detuning finds symmetric effects on damping of the $S'$ and $A'$ modes, with $\alpha_{S'}=\alpha_0+2\alpha_{sp}\:z^2$ and $\alpha_{A'}=\alpha_0+2\alpha_{sp}(1-z^2)$, respectively, recovering perfect symmetric and antisymmetric mode values for $z=0$.  We assume that the field splitting shown in Figure \ref{fig1} a) gives an accurate measure of $2\omega_{ex}/\gamma$, as supported by the MOKE results.  This value goes into the denominator of $z$.  We find a reasonable fit to the dependence of $S$ and $A$ damping on Ru thickness, implicit in the coupling.  For the highest coupling pionts, the damping values closely reach the low-$z$ limit, and we believe that the "giant" spin pumping result of $2\alpha_{sp}$ is evident here.   

We would like to point out next that it was not {\it a-priori} obvious that the Py/Ru/Py SAF would exhibit the observed damping.  Ru could behave in two limits in the context of spin pumping: either as a passive spin-sink layer, or as a ballistic transmission layer supporting transverse spin-current transmission from one Py layer to the other.  Our results show that Ru behaves as the latter in this thickness range.  The symmetric-mode damping of the SAF structure, extrapolated back to zero Ru thickness, is identical within experimental resolution ($\sim\textrm{10}^{-4}$) to that of a single Py film 16 nm thick measured in {\it nc}-FMR ("$\alpha_0$" line in Fig \ref{fig1}b.)  If Ru, or the Py/Ru interface, depolarized pumped spin current very strongly over this thickness range as has been proposed for Pt\cite{rojasPRL14}, we would expect an immediate increase in damping of the acoustic mode by the amount of $\sim\alpha_{sp}$.  Instead, the volume-dependent Ru depolarization in spin pumping has an (exponential) characteristic length of $\lambda_{SD}\sim\textrm{10 nm}$\cite{ghoshPRL12}, and attenuation over the range explored of $\sim$ 1 nm is negligible.

{\it Perspectives:}  Finally, we would like to highlight some implications of the study.  First, as the study confirms the prediction of a "giant" spin pumping effect as proposed by Takahashi\cite{takahashiAPL15}, it is plausible that the greatly enhanced values of spin accumulation predicted there may be supported by Ru in Py/Ru/Py synthetic antiferromagnets (SAFs).  These spin accumulations would differ strongly in the excitation of symmetric and antisymmetric modes, and may then provide a clear signature in time-resolved x-ray magnetooptical techniques\cite{baileyNComm13}, similar to the observation of static spin accumulation in Cu reported recently\cite{kukrejaPRL15}.  

Second, in most device applications of synthetic antiferromagnets, it is not desirable to excite the antisymmetric (optical) mode.  SAFs are used in the pinned layer of MTJ/spin valve structures to increase exchange bias and in the free layer to decrease (magnetostatic) stray fields.  Both of these functions are degraded if the optical, or asymmetric mode of the SAF is excited.  According to our results, at GHz frequencies near FMR, the susceptibility of the antisymmetric mode is reduced substantially, here by a factor of two (from 1/$\alpha$) for {\it nc-FMR}, due to spin pumping.  This reduction of $\chi$ on resonance will scale inversely with layer thickness.  The damping, and susceptibility, of the desired symmetric (acoustic) mode is unchanged, on the other hand, implying that spin pumping favors the excitation the symmetric mode for thin Ru, the typical operating point.

We acknowledge NSF-DMR-1411160 for support.

\printfigures

\bibliography{local}

\end{document}